\documentclass[aps,prb,twocolumn,superscriptaddress,showpacs]{revtex4}

\usepackage{graphicx}% Include figure files
\usepackage{amsmath}
\usepackage{amssymb}

\renewcommand{\vec}[1]{\boldsymbol{#1}}

\begin{document}

\title{Electronic structure of kinetic energy driven superconductors
in the presence of bilayer splitting}

\author{Yu Lan}
\affiliation{Department of Physics, Beijing Normal University,
Beijing 100875, China}

\author{Jihong Qin}
\affiliation{Department of Physics, Beijing University of Science
and Technology, Beijing 100083, China}

\author{Shiping Feng}
\affiliation{Department of Physics, Beijing Normal University,
Beijing 100875, China}

\date{\today}

\begin{abstract}
Within the framework of the kinetic energy driven superconductivity,
the electronic structure of bilayer cuprate superconductors in the
superconducting state is studied. It is shown that the electron
spectrum of bilayer cuprate superconductors is split into the
bonding and antibonding components by the bilayer splitting, then
the observed peak-dip-hump structure around the $[\pi,0]$ point is
mainly caused by this bilayer splitting, with the superconducting
peak being related to the antibonding component, and the hump being
formed by the bonding component. The spectral weight increases with
increasing the doping concentration. In analogy to the normal state
case, both electron antibonding peak and bonding hump have the weak
dispersions around the $[\pi,0]$ point.
\end{abstract}

\pacs{74.20.Mn, 74.20.-z, 74.25.Jb}

\maketitle

%%%%%%%%%%%%%%%%%%%%%%%%%%%%%%%%%%%%%%%%%%%%%%%%%

\section{Introduction}

The parent compounds of cuprate superconductors are the Mott
insulators with an antiferromagnetic (AF) long-range order (AFLRO),
then via the charge carrier doping, one can drive these materials
through a metal-insulating transition and enter the superconducting
(SC) dome \cite{kastner,shen,campuzano}. It has become clear in the
past twenty years that cuprate superconductors are among the most
complex systems studied in condensed matter physics
\cite{kastner,shen,campuzano}. The complications arise mainly from
(1) a layered crystal structure with one or more CuO$_{2}$ planes
per unit cell separated by insulating layers which leads to a
quasi-two-dimensional electronic structure, and (2) extreme
sensitivity of the physical properties to the compositions
(stoichiometry) which control the carrier density in the CuO$_{2}$
plane \cite{kastner,shen,campuzano}. As a consequence, both
experimental investigation and theoretical understanding are
extremely difficult.

By virtue of systematic studies using the angle-resolved
photoemission spectroscopy (ARPES), the low-energy electronic
structure of cuprate superconductors in the SC state is
well-established by now \cite{shen,campuzano}, where an agreement
has emerged that the electronic quasiparticle-like excitations are
well defined, and are the entities participating in the SC pairing.
In particular, the lowest energy states are located at the $[\pi,0]$
point of the Brillouin zone, where the d-wave SC gap function is
maximal, then the most contributions of the electron spectral
function come from the $[\pi,0]$ point \cite{shen,campuzano}.
Moreover, some ARPES experimental results unambiguously established
the Bogoliubov-quasiparticle nature of the sharp SC quasiparticle
peak near the $[\pi,0]$ point \cite{campuzano1,matsui}, then the SC
coherence of the quasiparticle peak is described by the simple
Bardeen-Cooper-Schrieffer (BCS) formalism \cite{bcs}. However, there
are numerous anomalies for different families of cuprate
superconductors, which complicate the physical properties of the
electronic structure \cite{shen,campuzano}. Among these anomalies is
the dramatic change in the spectral lineshape around the $[\pi,0]$
point first observed on the bilayer cuprate superconductor
Bi$_{2}$Sr$_{2}$CaCu$_{2}$O$_{8+\delta}$, where a sharp
quasiparticle peak develops at the lowest binding energy, followed
by a dip and a hump, giving rise to the so-called peak-dip-hump
(PDH) structure in the electron spectrum
\cite{dessau,randeria,fedorov}. Later, this PDH structure was also
found in YBa$_{2}$Cu$_{3}$O$_{7-\delta}$ \cite{lu} and in
Bi$_{2}$Sr$_{2}$Ca$_{2}$Cu$_{3}$O$_{10+\delta}$ \cite{sato}.
Furthermore, although the sharp quasiparticle peaks are identified
in the SC state along the entire Fermi surface, the PDH structure is
most strongly developed around the $[\pi,0]$ point
\cite{shen,dessau,randeria,fedorov,lu,sato}.

The appearance of the PDH structure in bilayer cuprate
superconductors in the SC state is the mostly remarkable effect,
however, its full understanding is still a challenging issue. The
earlier works \cite{campuzano2,shen} gave the main impetus for a
phenomenological description of the single-particle excitations in
terms of an interaction between quasiparticles and collective modes,
which is of fundamental relevance to the nature of superconductivity
and the pairing mechanism in cuprate superconductors. However, the
different interpretive scenario has been proposed
\cite{kordyuk,shen}. This followed from the observation of the
bilayer splitting (BS) for both normal and SC states in a wide
doping range \cite{dfeng,chuang,borisenko}. This BS of the CuO$_{2}$
plane derives the electronic structure in the bonding and
antibonding bands due to the present of CuO$_{2}$ bilayer blocks in
the unit cell of bilayer cuprate superconductors, then the main
features of the PDH structure is caused by the BS
\cite{kordyuk,dfeng,chuang,borisenko}, with the peak and hump
corresponding to the antibonding and bonding bands, respectively.
Furthermore, some ARPES experimental data measured above and below
the SC transition temperature show that this PDH structure is
totally unrelated to superconductivity \cite{dfeng}. The recent
ARPES experimental results reported by several groups support this
scenario, and most convincingly suggested that the PDH structure
originates from the BS at any doping levels \cite{dfeng1}. To the
best of our knowledge, this PDH structure in bilayer cuprate
superconductors has not been treated starting from a microscopic SC
theory.

Within the single layer $t$-$t'$-$J$ model, the electronic structure
of the single layer cuprate superconductors in the SC state has been
discussed \cite{guo} based on the framework of the kinetic energy
driven superconductivity \cite{feng}, and the main features of the
ARPES experiments on the single layer cuprate superconductors have
been reproduced, including the doping and temperature dependence of
the electron spectrum and quasiparticle dispersion. In this paper,
we study the electronic structure of bilayer cuprate superconductors
in the SC state along with this line. Within the kinetic energy
driven SC mechanism \cite{feng}, we employed the $t$-$t'$-$J$ model
by considering the bilayer interaction, and then show explicitly
that the BS occurs due to this bilayer interaction. In this case,
the electron spectrum is split into the bonding and antibonding
components by this BS, then the SC peak is closely related to the
antibonding component, while the hump is mainly formed by the
bonding component. In other words, the well pronounced PDH structure
in the electron spectrum of bilayer cuprate superconductors is
mainly caused by the BS. Furthermore, the spectral weight in the
$[\pi,0]$ point increases with increasing the doping concentration.
In analogy to the normal-state case \cite{dfeng,kordyuk1,mori,lan},
both electron antibonding peak and bonding hump have the weak
dispersions around the $[\pi,0]$ point, in qualitative agreement
with the experimental observation on bilayer cuprate superconductors
in the SC state \cite{shen,dessau,randeria,fedorov,lu,sato}.

The paper is organized as follows. The basic formalism is presented
in Sec. II, where we generalize the kinetic energy driven
superconductivity from the previous single layer case
\cite{feng,guo} to the bilayer case, and then evaluate explicitly
the longitudinal and transverse components of the electron normal
and anomalous Green's functions (hence the bonding and antibonding
electron spectral functions). Within this theoretical framework, we
discuss the electronic structure of bilayer cuprate superconductors
in the SC state in Sec. III. It is shown that the striking PDH
structure in bilayer cuprate superconductors is closely related to
the BS. Finally, we give a summary and discussions in Sec. IV.

\section{Formalism}

It has been shown from the ARPES experiments \cite{shen,kim} that
the two-dimensional $t$-$t'$-$J$ model is of particular relevance to
the low energy features of cuprate superconductors. For discussions
of the physical properties of bilayer cuprate superconductors, the
$t$-$t'$-$J$ model can be expressed by including the bilayer
interactions as,
\begin{eqnarray}
H&=&-t\sum_{i\hat{\eta}a\sigma}C^{\dagger}_{ia\sigma}C_{i+\hat{\eta}a\sigma}
+t'\sum_{i\hat{\tau}a\sigma}C^{\dagger}_{ia\sigma}C_{i+\hat{\tau}a\sigma}\nonumber\\
&-&\sum_{i\sigma}t_{\perp}(i)(C^{\dagger}_{i1\sigma}C_{i2\sigma}+H.c.)
+\mu\sum_{ia\sigma}C^{\dagger}_{ia\sigma}C_{ia\sigma}\nonumber\\
&+&J\sum_{i\hat{\eta}a}{\bf S}_{ia} \cdot {\bf
S}_{i+\hat{\eta}a}+J_{\perp}\sum_{i}{\bf S}_{i1} \cdot {\bf
S}_{i2},\label{ttjmodel}
\end{eqnarray}
supplemented by an important on-site local constraint $\sum_\sigma
C_{ia\sigma}^{\dagger}C_{ia\sigma}\leq 1$ to avoid the double
occupancy, where $\hat{\eta}=\pm\hat{x},\pm\hat{y}$ representing the
nearest neighbors of a given site $i$, $\hat{\tau}=\pm \hat{x}\pm
\hat{y}$ representing the next nearest neighbors of a given site
$i$, $a=1,2$ is plane index, $C^{\dagger}_{ia\sigma}$
($C_{ia\sigma}$) is the electron creation (annihilation) operator,
${\bf S}_{ia}=C^{\dagger}_{ia}{\vec\sigma}C_{ia}/2$ is the spin
operator with the Pauli matrices ${\vec\sigma} =(\sigma_{x},
\sigma_{y},\sigma_{z})$, $\mu$ is the chemical potential, and the
interlayer coherent hopping has the form,
\begin{eqnarray}
t_{\perp}({\bf k})={t_{\perp}\over 4}(\cos k_{x} -\cos k_{y})^{2},
\end{eqnarray}
which is strongly anisotropic and follows the theoretical
predictions \cite{chakarvarty}. In particular, this momentum
dependent form (2) has been experimentally verified
\cite{dfeng,chuang}. For this $t$-$t'$-$J$ model (1), it has been
argued that crucial requirement is to impose the electron single
occupancy local constraint for a proper understanding of the
physical properties of cuprate superconductors. To incorporate the
electron single occupancy local constraint, the charge-spin
separation (CSS) fermion-spin theory has been proposed \cite{feng1},
where the constrained electron operators are decoupled as,
$C_{ia\uparrow}=h^{\dagger}_{ia\uparrow} S^{-}_{ia}$ and
$C_{ia\downarrow}=h^{\dagger}_{ia\downarrow} S^{+}_{ia}$, with the
spinful fermion operator $h_{ia\sigma}= e^{-i\Phi_{ia\sigma}}
h_{ia}$ represents the charge degree of freedom together with some
effects of the spin configuration rearrangements due to the presence
of the doped hole itself (dressed holon), while the spin operator
$S_{ia}$ represents the spin degree of freedom, then the bilayer
$t$-$t'$-$J$ Hamiltonian (1) can be expressed in this CSS
fermion-spin representation as,
\begin{eqnarray}\label{csshamiltion}
H&=&t\sum_{i\hat{\eta}a}(h^{\dagger}_{i+\hat{\eta}a\uparrow}
h_{ia\uparrow}S^{+}_{ia}S^{-}_{i+\hat{\eta}a}+
h^{\dagger}_{i+\hat{\eta}a\downarrow}h_{ia\downarrow}S^{-}_{ia}
S^{+}_{i+\hat{\eta}a})\nonumber\\
&-&t'\sum_{i\hat{\tau}a} (h^{\dagger}_{i+\hat{\tau}a\uparrow}
h_{ia\uparrow}S^{+}_{ia}S^{-}_{i+\hat{\tau}a}+h^{\dagger}_{i+\hat{\tau}a\downarrow}h_{ia\downarrow}
S^{-}_{ia}S^{+}_{i+\hat{\tau}a})\nonumber\\
&+&\sum_{i}t_{\perp}(i)
(h^{\dagger}_{i2\uparrow}h_{i1\uparrow}S^{+}_{i1}S^{-}_{i2}+
h^{\dagger}_{i1\uparrow}h_{i2\uparrow}S^{+}_{i2}S^{-}_{i1}\nonumber\\
&+&h^{\dagger}_{i2\downarrow}h_{i1\downarrow}S^{-}_{i1} S^{+}_{i2}
+h^{\dagger}_{i1\downarrow}h_{i2\downarrow}S^{-}_{i2}
S^{+}_{i1})-\mu\sum_{ia\sigma}h^{\dagger}_{ia\sigma}h_{ia\sigma}\nonumber\\
&+&{J_{\rm eff}}\sum_{i\hat{\eta}a}{\bf S}_{ia}\cdot {\bf
S}_{i+\hat{\eta}a}+ {J_{\rm eff\perp}}\sum_{i}{\bf S}_{i1}\cdot {\bf
S}_{i2},
\end{eqnarray}
where $J_{\rm eff}=J(1-\delta)^{2}$, $J_{\rm eff\perp}=J_{\perp}
(1-\delta)^{2}$, and $\delta=\langle h^{\dagger}_{ia\sigma}
h_{ia\sigma}\rangle=\langle h^{\dagger}_{ia} h_{ia}\rangle$ is the
doping concentration. It has been shown that the electron single
occupancy local constraint is satisfied in analytical calculations
within this CSS fermion-spin theory, and the double spinful fermion
occupancy are ruled out automatically \cite{feng1}. Although in
common sense $h_{ia\sigma}$ is not a real spinful fermion, it
behaves like a spinful fermion \cite{feng1}. As in the single layer
case \cite{guo}, the kinetic energy terms in the bilayer
$t$-$t'$-$J$ model have been transferred as the dressed holon-spin
interactions, which can induce the dressed holon pairing state
(hence the electron Cooper pairing state) by exchanging spin
excitations in the higher power of the doping concentration. Before
calculation of the electron normal and anomalous Green's functions
of the bilayer system in the SC state, we firstly introduce the SC
order parameter. As we have mentioned above, there are two coupled
CuO$_{2}$ planes in the unit cell, and in this case, the SC order
parameters for the electron Cooper pair is a matrix
$\Delta=\Delta_{L}+\sigma_{x} \Delta_{T}$, with the longitudinal and
transverse SC order parameters in the CSS fermion-spin theory can be
expressed as,
\begin{subequations}
\begin{eqnarray}
\Delta_{L}&=&\langle C^{\dagger}_{ia\uparrow}
C^{\dagger}_{i+\hat{\eta}a\downarrow}-C^{\dagger}_{ia\downarrow}
C^{\dagger}_{i+\hat{\eta}a\uparrow}\rangle\nonumber\\
&=&\langle h_{ia\uparrow}
h_{i+\hat{\eta}a\downarrow}S^{+}_{ia}S^{-}_{i+\hat{\eta}a}-
h_{ia\downarrow}h_{i+\hat{\eta}a\uparrow}S^{-}_{ia}
S^{+}_{i+\hat{\eta}a}\rangle\nonumber\\
&=&-\chi_{1}\Delta_{hL},\\
\Delta_{T}&=&\langle C^{\dagger}_{i1\uparrow}
C^{\dagger}_{i2\downarrow}- C^{\dagger}_{i1\downarrow}
C^{\dagger}_{i2\uparrow}\rangle \nonumber\\
&=&\langle h_{i1\uparrow}
h_{i2\downarrow}S^{+}_{i1}S^{-}_{i2}-h_{i1\downarrow}
h_{i2\uparrow}S^{-}_{i1}S^{+}_{i2}\rangle\nonumber\\
&=&-\chi_{\perp}\Delta_{hT},
\end{eqnarray}
\end{subequations}
respectively, where the spin correlation functions $\chi_{1}=\langle
S_{ia}^{+}S_{i+\hat{\eta}a}^{-}\rangle$ and $\chi_{\perp}=\langle
S^{+}_{i1}S^{-}_{i2}\rangle$, and the longitudinal and transverse
dressed holon pairing order parameters $\Delta_{hL}=\langle
h_{i+\hat{\eta}a\downarrow}h_{ia\uparrow}-h_{i+\hat{\eta}a\uparrow}
h_{ia\downarrow}\rangle$ and $\Delta_{hT}=\langle h_{i2\downarrow}
h_{i1\uparrow}-h_{i2\uparrow} h_{i1\downarrow}\rangle$.

Within the $t$-$J$ type model, robust indications of
superconductivity with the d-wave symmetry in doped cuprates have
been found by using numerical techniques \cite{sorella}. On the
other hand, it has been argued that the SC transition in doped
cuprates is determined by the need to reduce the frustrated kinetic
energy \cite{anderson}. Although it is not necessary for the strong
coupling of the electron quasiparticles and a pairing boson in their
arguments \cite{anderson}, a series of the inelastic neutron
scattering experimental results provide a clear link between the
electron quasiparticles and magnetic excitations \cite{dai,arai}. In
particular, an impurity-substitution effect on the low energy
dynamics has been studied by virtue of the ARPES measurement
\cite{terashima}, this impurity-substitution effect is a magnetic
analogue of the isotope effect used for the conventional
superconductors. These experimental results \cite{terashima} reveal
that the impurity-induced changes in the electron self-energy show a
good correspondence to those of the magnetic excitations, indicating
the importance of the magnetic fluctuation to the electron pairing
in cuprate superconductors. Recently, we \cite{feng} have developed
the kinetic energy driven SC mechanism based on the CSS fermion-spin
theory \cite{feng1}, where the dressed holons interact occurring
directly through the kinetic energy by exchanging spin excitations,
leading to a net attractive force between dressed holons, then the
electron Cooper pairs originating from the dressed holon pairing
state are due to the charge-spin recombination, and their
condensation reveals the SC ground-state. Within this SC mechanism
\cite{feng}, the doping and temperature dependence of the electron
spectral function of the single layer cuprate superconductors in the
SC state has been discussed \cite{guo}. In this section, our main
goal is to generalize these analytical calculations from the single
layer case to the bilayer system. As in the case for the SC order
parameter, the full dressed holon normal and anomalous Green's
functions can also be expressed as $g({\bf k},\omega)=g_{L}({\bf
k},\omega)+ \sigma_{x}g_{T}({\bf k},\omega)$ and $\Im^{\dagger}({\bf
k},\omega) =\Im^{\dagger}_{L}({\bf k},\omega) +
\sigma_{x}\Im^{\dagger}_{L} ({\bf k},\omega)$, respectively. We now
can follow the previous discussions for the single layer case
\cite{guo,feng}, and evaluate explicitly these corresponding
longitudinal and transverse parts of the full dressed holon normal
and anomalous Green's functions as [see the Appendix],
\begin{widetext}
\begin{subequations}
\begin{eqnarray}
g_{L}({\bf k},\omega) &=& {1\over 2}\sum_{\nu=1,2}Z^{(\nu)}_{hFA}
\left ({U^{2}_{h\nu{\bf k}}\over\omega-E_{h\nu{\bf k}}}+
{V^{2}_{h\nu{\bf k}}\over\omega+E_{h\nu{\bf k}}}\right), \\
g_{T}({\bf k},\omega)&=&{1\over 2}\sum_{\nu=1,2}(-1)^{\nu+1}
Z^{(\nu)}_{hFA}\left ({U^{2}_{h\nu{\bf k}}\over\omega- E_{h\nu{\bf
k}}}+{V^{2}_{h\nu{\bf k}}\over\omega+E_{h\nu{\bf k}}}
\right ), \\
\Im^{\dagger}_{L}({\bf k},\omega)&=&-{1\over 2}\sum_{\nu=1,2}
Z^{(\nu)}_{hFA}{\bar{\Delta}_{hz}^{(\nu)}({\bf k})\over 2E_{h\nu{\bf
k}}}\left ({1\over \omega-E_{h\nu{\bf k}}}-{1\over
\omega+ E_{h\nu{\bf k}}}\right ), \\
\Im^{\dagger}_{T}({\bf k},\omega)&=&-{1\over 2}\sum_{\nu=1,2}
(-1)^{\nu+1}Z^{(\nu)}_{hFA}{\bar{\Delta}_{hz}^{(\nu)}({\bf k})\over
2E_{h\nu{\bf k}}}\left ({1\over\omega-E_{h\nu{\bf k}}}-{1\over
\omega+E_{h\nu{\bf k}}}\right ),
\end{eqnarray}
\end{subequations}
\end{widetext}
where the dressed holon quasiparticle coherence
factors $U^{2}_{h\nu{\bf k}}=[1+\bar{\xi}_{\nu{\bf k}}/ E_{h\nu{\bf
k}}]/2$ and $V^{2}_{h\nu{\bf k}}=[1-\bar{\xi}_{\nu{\bf
k}}/E_{h\nu{\bf k}}] /2$, the dressed holon quasiparticle dispersion
$E_{h\nu{\bf k}}= \sqrt{[{\bar\xi}_{\nu{\bf k}}]^{2}+
\mid\bar{\Delta}_{hz}^{(\nu)} ({\bf k})\mid^{2}}$, the renormalized
dressed holon excitation spectrum ${\bar\xi}_{\nu{\bf
k}}=Z^{(\nu)}_{hFA}\xi_{\nu{\bf k}}$, with the mean-field (MF)
dressed holon excitation spectrum $\xi_{\nu{\bf
k}}=Zt\chi_{1}\gamma_{\bf{k}}-Zt'\chi_{2}
\gamma{'}_{\bf{k}}-\mu+(-1)^{\nu+1}\chi_{\perp}t_{\perp}({\bf{k}})$,
where the spin correlation function $\chi_{2}=\langle S_{ia}^{+}
S_{i+\hat{\tau}a}^{-}\rangle$, $\gamma_{\bf{k}}=(1/Z)
\sum_{\hat{\eta}}{e^{i{\bf{k}}\cdot{\hat{\eta}}}}$,
$\gamma{'}_{\bf{k}}=(1/Z) \sum_{\hat{\tau}}{e^{i{\bf{k}}
\cdot{\hat{\tau}}}}$, $Z$ is the number of the nearest neighbor or
next nearest neighbor sites, the renormalized dressed holon pair gap
function $\bar{\Delta}_{hz}^{(\nu)}({\bf k})=Z^{(\nu)}_{hFA}
[\bar{\Delta}_{hL}({\bf k})+(-1)^{\nu+1}\bar{\Delta}_{hT}({\bf k})
]$, with $\nu=1$ ( $\nu=2$) for the bonding (antibonding) case,
where $\bar{\Delta}_{hL}({\bf k})=\Sigma^{(h)}_{2L}({\bf k},\omega)
\mid_{\omega=0}=\bar{\Delta}_{hL}\gamma^{(d)}_{\bf k}$, with
$\gamma^{(d)}_{\bf k}=({\rm cos}k_{x}-{\rm cos} k_{y})/2$,
$\bar{\Delta}_{hT}({\bf k})=\Sigma^{(h)}_{2T} ({\bf k},\omega)
\mid_{\omega=0}=\bar{\Delta}_{hT}$, the dressed holon quasiparticle
coherent weights $Z^{(1)-1}_{hFA}=Z^{-1}_{hF1}-Z^{-1}_{hF2}$,
$Z^{(2)-1}_{hFA}=Z^{-1}_{hF1}+Z^{-1}_{hF2}$, with $Z^{-1}_{hF1}=1-
\Sigma^{(ho)}_{1L} ({\bf k}_{0},\omega)\mid_{\omega=0}$, and
$Z^{-1}_{hF2}=\Sigma^{(ho)}_{1T}({\bf k}_{0},\omega)\mid_{\omega=0}$
, where ${\bf k}_{0}=[\pi,0]$, $\Sigma^{(ho)}_{1L}({\bf k},\omega)$
and $\Sigma^{(ho)}_{1T}({\bf k} ,\omega)$ are the corresponding
antisymmetric parts of the longitudinal and transverse dressed holon
self-energy functions $\Sigma^{(h)}_{1L}({\bf k},\omega)$ and
$\Sigma^{(h)}_{1T}({\bf k},\omega)$, while the longitudinal and
transverse parts of the dressed holon self-energy functions
$\Sigma^{(h)}_{1}({\bf k},\omega)$ and $\Sigma^{(h)}_{2}({\bf
k},\omega)$ have been evaluated as,
\begin{widetext}
\begin{subequations}
\begin{eqnarray}
\Sigma^{(h)}_{1L}({\bf k},i\omega_{n})&=&{1\over N^{2}}\sum_{\bf p,
q}[ R^{(1)}_{\bf{p+q+k}}{1\over\beta}\sum_{ip_{m}}g_{L} ({\bf p+k},
ip_{m}+i\omega_{n})\Pi_{LL}({\bf p},{\bf q},ip_{m})\nonumber \\
&+&R^{(2)}_{\bf{p+q+k}}{1\over\beta}\sum_{ip_{m}}g_{T}({\bf p+k},
ip_{m}+i\omega_{n})\Pi_{TL}({\bf p},{\bf q},ip_{m})],\\
\Sigma^{(h)}_{1T}({\bf k},i\omega_{n})&=&{1\over N^{2}}\sum_{\bf p,
q}[R^{(1)}_{\bf{p+q+k}}{1\over\beta}\sum_{ip_{m}} g_{T}({\bf p+k},
ip_{m}+i\omega_{n})\Pi_{TT}({\bf p},{\bf q},ip_{m})\nonumber\\
&+&R^{(2)}_{\bf{p+q+k}}{1\over\beta}\sum_{ip_{m}}g_{L}({\bf p+k},
ip_{m}+i\omega_{n})\Pi_{LT}({\bf p},{\bf q},ip_{m})],\\
\Sigma^{(h)}_{2L}({\bf k},i\omega_{n})&=&{1\over N^{2}}\sum_{\bf p,
q}[R^{(1)}_{\bf{p+q+k}}{1\over \beta}\sum_{ip_{m}} \Im^{\dag}_{L}
(-{\bf p-k},-ip_{m}-i\omega_{n})\Pi_{LL}({\bf p},{\bf q},ip_{m})
\nonumber \\
&+& R^{(2)}_{\bf{p+q+k}}{1\over\beta}\sum_{ip_{m}}\Im^{\dag}_{T}
(-{\bf p-k},-ip_{m}-i\omega_{n})\Pi_{TL}({\bf p},{\bf q},ip_{m})],
\\
\Sigma^{(h)}_{2T}({\bf k},i\omega_{n})&=&{1\over N^{2}}\sum_{\bf p,
q} [R^{(1)}_{\bf{p+q+k}}{1\over \beta}\sum_{ip_{m}} \Im^{\dag}_{T}
(-{\bf p-k},-ip_{m}-i\omega_{n})\Pi_{TT}({\bf p},{\bf q},ip_{m})
\nonumber \\
&+&R^{(2)}_{\bf{p+q+k}}{1\over \beta}\sum_{ip_{m}} \Im^{\dag}_{L}
(-{\bf p-k},-ip_{m}-i\omega_{n})\Pi_{LT}({\bf p},{\bf q}, ip_{m})],
\end{eqnarray}
\end{subequations}
\end{widetext}
where $R^{(1)}_{\bf k}=[Z(t\gamma_{\bf
k}-t'\gamma'_{\bf k})]^{2}+ t_{\perp}^{2}({\bf k})$, $R^{(2)}_{\bf
k}=2Z(t\gamma_{\bf k}- t' \gamma'_{\bf k})t_{\perp}({\bf k})$, and
the spin bubbles $\Pi_{\eta,\eta'}({\bf p},{\bf q},ip_{m})=
(1/\beta)\sum_{iq_{m}} D^{(0)}_{\eta}({\bf q},iq_{m})
D^{(0)}_{\eta'}({\bf q+p},iq_{m}+ ip_{m})$, with $\eta=L,T$ and
$\eta'=L,T$, and the MF spin Green's function $D^{(0)}({\bf
k},\omega)=D^{(0)}_{L}({\bf k},\omega) + \sigma_{x}D^{(0)}_{T}({\bf
k},\omega)$, with the corresponding longitudinal and transverse
parts have been given by \cite{lan},
\begin{subequations}
\begin{eqnarray}
D^{(0)}_{L}({\bf k},\omega)&=&{1\over 2}\sum_{\nu=1,2}
{B_{\nu{\bf k}}\over \omega^{2}-\omega^{2}_{\nu{\bf k}}}, \\
D^{(0)}_{T}({\bf k},\omega)&=&{1\over 2}
\sum_{\nu=1,2}(-1)^{\nu+1}{B_{\nu{\bf k}}\over \omega^{2}-
\omega^{2}_{\nu{\bf k}}},
\end{eqnarray}
\end{subequations}
where $B_{\nu{\bf k}}=\lambda(A_{1}\gamma_{\bf k}-A_{2})-\lambda{'}
(2\chi_{2}^{z}\gamma {'}_{\bf k}-\chi_{2})-J_{\rm{eff}\perp}
[\chi_{\perp}+2\chi_{\perp}^{z}(-1)^{\nu}][\epsilon_{\perp}({\bf k}
)+(-1)^{\nu}]$, $A_{1}=2\epsilon_{\parallel}\chi_{1}^{z}+\chi_{1}$,
$A_{2}=\epsilon_{\parallel}\chi_{1}+2\chi_{1}^{z}$, $\lambda=
2ZJ_{\rm eff}$, $\lambda{'}=4Z\phi_{2}t'$, $\epsilon_{\parallel}=
1+2t\phi_{1}/J_{\rm eff}$, $\epsilon_{\perp}({\bf{k}})=1+
4\phi_{\perp}t_{\perp}({\bf{k}})/J_{\rm eff\perp}$, the spin
correlation functions $\chi_{1}^{z}=\langle S_{ia}^{z}
S_{i+\hat{\eta}a}^{z}\rangle$, $\chi_{2}^{z}=\langle S_{ia}^{z}
S_{i+\hat{\tau}a}^{z}\rangle$,  $\chi^{z}_{\perp}=\langle S_{i1}^{z}
S_{i2}^{z}\rangle$, the dressed holon particle-hole order parameters
$\phi_{1}=\langle h^{\dagger}_{ia\sigma} h_{i+\hat{\eta}a\sigma}
\rangle$, $\phi_{2}=\langle h^{\dagger}_{ia\sigma}
h_{i+\hat{\tau}a\sigma}\rangle$, $\phi_{\perp}=\langle
h^{\dagger}_{i1\sigma}h_{i2\sigma}\rangle$, and the MF spin
excitation spectrum,
\begin{widetext}
\begin{eqnarray}
\omega^{2}_{\nu{\bf k}}&=&\lambda^{2}\left [\left (A_{4}-\alpha
\epsilon_{\parallel}\chi_{1}^{z}\gamma_{\bf k}-{1\over 2Z}\alpha
\epsilon_{\parallel}\chi_{1}\right )(1-\epsilon_{\parallel}
\gamma_{\bf k})+{1\over 2}\epsilon_{\parallel}\left (A_{3}-{2\over
Z}\alpha\chi_{1}^{z} -\alpha\chi_{1}\gamma_{\bf k}\right )
(\epsilon_{\parallel}-\gamma_{\bf k})\right ]
\nonumber\\
&+&\lambda{'}^{2}\left [\alpha\left (\chi_{2}^{z}\gamma{'}_{\bf k}
-{Z-1\over 2Z}\chi_{2}\right )\gamma{'}_{\bf k}+{1\over 2}\left
(A_{5}-{2\over Z}\alpha\chi_{2}^{z}\right)\right]+\lambda
\lambda{'}\alpha \left[\chi_{1}^{z}(1-\epsilon_{\parallel}
\gamma_{\bf k})\gamma{'}_{\bf k}+{1\over 2}(\chi_{1}\gamma{'}_{\bf
k}-C_{2}) (\epsilon_{\parallel} -\gamma_{\bf
k})\right.\nonumber\\
&+&\left.\gamma{'}_{\bf k}(C_{2}^{z} -\epsilon_{\parallel}
\chi_{2}^{z}\gamma_{\bf k})-{1 \over 2}
\epsilon_{\parallel}(C_{2}-\chi_{2}\gamma_{\bf k}) \right] +\lambda
J_{\rm eff \perp}\alpha\left \{{1\over 2}\epsilon_{\perp} ({\bf
k})(\epsilon_{\parallel}-\gamma_{\bf k})[C_{\perp}+
\chi_{1}(-1)^{\nu}]\right .\nonumber\\
&+&\left.(1-\epsilon_{\parallel}\gamma_{\bf k})
[C_{\perp}^{z}+\chi_{1}^{z}\epsilon_{\perp}({\bf k})(-1)^{\nu}]
+[\epsilon_{\perp}({\bf k})+(-1)^{\nu}]\left [{1\over 2}
\epsilon_{\parallel}(C_{\perp}-\chi_{\perp}\gamma_{\bf k})+
(C_{\perp}^{z}-\epsilon_{\parallel}\chi_{\perp}^{z}\gamma_{\bf k})
(-1)^{\nu}\right ]\right \}\nonumber\\
&+&\lambda{'}J_{\rm eff \perp}\alpha\left \{\gamma{'}_{\bf k}
[C{'}_{\perp}^{z}+\chi_{2}^{z}\epsilon_{\perp}({\bf k})(-1)^{\nu}]
-{1 \over 2}\epsilon_{\perp}({\bf k})[C'_{\perp}+\chi_{2}(-1)^{\nu}]
\right . \nonumber\\
&+&\left . \left [{1 \over 2}(\chi_{\perp}\gamma{'}_{\bf k}-
C'_{\perp})+\chi_{\perp}^{z}\gamma{'}_{\bf k}(-1)^{\nu}\right ]
[\epsilon_{\perp}({\bf k})+(-1)^{\nu}]\right \}+{1\over 4}
J_{\rm{eff}\perp}^{2}[\epsilon_{\perp} ({\bf k})+(-1)^{\nu}]^{2},
\end{eqnarray}
\end{widetext}
where $A_{3}=\alpha C_{1}+(1-\alpha)/2Z$, $A_{4}=\alpha C_{1}^{z}+
(1-\alpha)/4Z$, $A_{5}=\alpha C_{3}+(1-\alpha)/2Z$, and the spin
correlation functions $C_{1}=(1/Z^{2})\sum_{\hat{\eta}\hat{\eta'}}
\langle S_{i+\hat{\eta}a}^{+}S_{i+\hat{\eta'}a}^{-}\rangle$, $C_{2}=
(1/Z^{2})\sum_{\hat{\eta} \hat{\tau}}\langle S_{i+\hat{\eta}a}^{+}
S_{i+\hat{\tau}a}^{-}\rangle$, $C_{3}=(1/Z^{2})
\sum_{\hat{\tau}\hat{\tau'}}\langle S_{i+\hat{\tau}a}^{+}
S_{i+\hat{\tau'}a}^{-}\rangle$, $C_{1}^{z}= (1/Z^{2})
\sum_{\hat{\eta}\hat{\eta'}}\langle S_{i+\hat{\eta}a}^{z}
S_{i+\hat{\eta'}a}^{z}\rangle$, $C_{2}^{z}=(1/Z^{2})
\sum_{\hat{\eta}\hat{\tau}}\langle S_{i+\hat{\eta}a}^{z}
S_{i+\hat{\tau}a}^{z}\rangle$, $C_{\perp}=(1/Z)\sum_{\hat{\eta}}
\langle S_{i1}^{+} S_{i+\hat{\eta}2}^{-}\rangle$, $C{'}_{\perp}=
(1/Z)\sum_{\hat{\tau}} \langle S_{i1}^{+} S_{i+\hat{\tau}2}^{-}
\rangle$, $C_{\perp}^{z}=(1/Z) \sum_{\hat{\eta}}\langle S_{i1}^{z}
S_{i+ \hat{\eta}2}^{z}\rangle$, and $C{'}_{\perp}^{z}=(1/Z)
\sum_{\hat{\tau}}\langle S_{i1}^{z}S_{i+ \hat{\tau}2}^{z}\rangle$.
In order to satisfy the sum rule of the spin correlation function
$\langle S^{+}_{ia}S^{-}_{ia}\rangle=1/2$ in the case without AFLRO,
the important decoupling parameter $\alpha$ has been introduced in
the above calculation as in the single layer case
\cite{guo,feng,lan}, which can be regarded as the vertex correction.

With the help of the longitudinal and transverse parts of the full
dressed holon normal and anomalous Green's functions in Eq. (5) and
MF spin Green's function in Eq. (7), we now can calculate the
electron normal and anomalous Green's functions $G(i-j,t-t')=
\langle\langle C_{i\sigma}(t);C^{\dagger}_{j\sigma}(t')\rangle
\rangle=G_{L}(i-j,t-t')+\sigma_{x} G_{T}(i-j,t-t')$ and
$\Gamma^{\dagger}(i-j,t-t')=\langle\langle C^{\dagger}_{i\uparrow}
(t);C^{\dagger}_{j\downarrow}(t')\rangle\rangle=
\Gamma^{\dagger}_{L}(i-j,t-t')+\sigma_{x}\Gamma^{\dagger}_{T}
(i-j,t-t')$, where these longitudinal and transverse parts are the
convolutions of the corresponding longitudinal and transverse parts
of the full dressed holon normal and anomalous Green's functions and
MF spin Green's function in the CSS fermion-spin theory, and can be
evaluated explicitly as,
\begin{widetext}
\begin{subequations}
\begin{eqnarray}
G_{L}({\bf k},\omega)&=&{1\over 8N}\sum_{\bf p}\sum_{\mu\nu}
Z_{hFA}^{(\mu)}{B_{\nu{\bf p}}\over\omega_{\nu{\bf p}}}\left [
L^{(1)}_{\mu\nu}({\bf k,p}) \left ({U^{2}_{h\mu{\bf p-k}}\over
\omega+E_{h\mu{\bf p-k}}-\omega_{\nu{\bf p}}}+{V^{2}_{h\mu{\bf p-k}}
\over \omega-E_{h\mu{\bf p-k}}+\omega_{\nu{\bf p}}}\right )\right .
\nonumber\\
&+&\left . L^{(2)}_{\mu\nu}({\bf k,p})\left ({U^{2}_{h\mu{\bf p-k}}
\over\omega+E_{h\mu{\bf p-k}}+\omega_{\nu{\bf p}}}+{V^{2}_{h\mu{\bf
p-k}}\over\omega-E_{h\mu{\bf p-k}}-\omega_{\nu{\bf p}}}\right )
\right ] ,\\
G_{T}({\bf k},\omega)&=&{1 \over 8N}\sum_{\bf p}\sum_{\mu\nu}
(-1)^{\mu+\nu}Z_{hFA}^{(\mu)}{B_{\nu{\bf p}}\over\omega_{\nu{\bf p}}
}\left [ L^{(1)}_{\mu\nu}({\bf k,p}) \left ({U^{2}_{h\mu{\bf p-k}}
\over\omega+E_{h\mu{\bf p-k}}-\omega_{\nu{\bf p}}}+{V^{2}_{h\mu{\bf
p-k}}\over \omega-E_{h\mu{\bf p-k}}+\omega_{\nu{\bf p}}}\right )
\right . \nonumber\\
&+&\left . L^{(2)}_{\mu\nu}({\bf k,p})\left ({U^{2}_{h\mu{\bf p-k}}
\over\omega+E_{h\mu{\bf p-k}}+\omega_{\nu{\bf p}}}+{V^{2}_{h\mu{\bf
p-k}}\over\omega-E_{h\mu{\bf p-k}}-\omega_{\nu{\bf p}}}\right )
\right ] ,\\
\Gamma^{\dagger}_{L}({\bf k},\omega)&=&{1\over 8N}\sum_{\bf p}
\sum_{\mu\nu}Z^{(\mu)}_{hFA}{\bar{\Delta}_{hz}^{(\mu)}({\bf p-k})
\over 2E_{h\mu{\bf p-k}}}{B_{\nu{\bf p}}\over\omega_{\nu{\bf p}}}
\left [ L^{(1)}_{\mu\nu}({\bf k,p})\left ({1\over \omega-
E_{h\mu{\bf p-k}}+\omega_{\nu{\bf p}}}\right .\right .\nonumber \\
&-&\left.{1\over\omega+E_{h\mu{\bf p-k}}-\omega_{\nu{\bf p}}}\right)
+\left .L^{(2)}_{\mu\nu}({\bf k,p})\left ({1\over\omega- E_{h\mu{\bf
p-k}}-\omega_{\nu{\bf p}}}-{1\over\omega+E_{h\mu{\bf
p-k}}+\omega_{\nu{\bf p}}}\right )\right ], \\
\Gamma^{\dagger}_{T}({\bf k},\omega)&=&{1\over 8N}\sum_{\bf p}
\sum_{\mu\nu}(-1)^{\mu+\nu}Z^{(\mu)}_{hFA}{\bar{\Delta}_{hz}^{(\mu)}
({\bf p-k})\over 2E_{h\mu{\bf p-k}}}{B_{\nu{\bf p}}\over
\omega_{\nu{\bf p}}}\left [ L^{(1)}_{\mu\nu}({\bf k,p})\left ({1
\over \omega- E_{h\mu{\bf p-k}}+\omega_{\nu{\bf p}}}\right .\right .
\nonumber \\
&-&\left .{1\over\omega+E_{h\mu{\bf p-k}}-\omega_{\nu{\bf p}}}
\right) + \left . L^{(2)}_{\mu\nu}({\bf k,p})\left ({1\over\omega-
E_{h\mu{\bf p-k}} -\omega_{\nu{\bf p}}}- {1\over\omega+E_{h\mu{\bf
p-k}}+ \omega_{\nu{\bf p}}}\right )\right ],
\end{eqnarray}
\end{subequations}
\end{widetext}
where $L^{(1)}_{\mu\nu}({\bf k,p}) = [{\rm coth}
(\beta\omega_{\nu{\bf p}}/2)-{\rm th}(\beta E_{h\mu{\bf p-k}}/2)]/2$
and $L^{(2)}_{\mu\nu}({\bf k,p})=[{\rm coth}(\beta\omega_{\nu{\bf p}
}/2)+{\rm th}(\beta E_{h\mu{\bf p-k}}/2)]/2$, then the longitudinal
and transverse parts of the electron spectral function $A_{L}({\bf
k},\omega)=-2{\rm Im}G_{L}({\bf k},\omega)$ and $A_{T}({\bf
k},\omega)=-2{\rm Im}G_{T}({\bf k},\omega)$ and SC gap function
$\Delta_{L}({\bf k})=(1/\beta)\sum_{i\omega_{n}}
\Gamma^{\dagger}_{L}({\bf k},i\omega_{n} )$ and $\Delta_{T}({\bf k})
=(1/\beta)\sum_{i\omega_{n}}\Gamma^{\dagger}_{T}({\bf k},
i\omega_{n})$ are obtained as,
\begin{widetext}
\begin{subequations}
\begin{eqnarray}
A_{L}({\bf k},\omega)&=&\pi{1\over 4N}\sum_{\bf p}\sum_{\mu\nu}
Z_{hFA}^{(\mu)}{B_{\nu{\bf p}}\over\omega_{\nu{\bf p}}}\{
L^{(1)}_{\mu\nu}({\bf k,p})[U^{2}_{h\mu{\bf p-k}}
\delta(\omega+E_{h\mu{\bf p-k}}-\omega_{\nu{\bf p}})+V^{2}_{h\mu{\bf
p-k}}\delta(\omega-E_{h\mu{\bf p-k}}+ \omega_{\nu{\bf p}})]\nonumber\\
&+&L^{(2)}_{\mu\nu}({\bf k,p})[U^{2}_{h\mu{\bf p-k}}
\delta(\omega+E_{h\mu{\bf p-k}}+ \omega_{\nu{\bf p}})
+V^{2}_{h\mu{\bf p-k}}\delta(\omega-E_{h\mu{\bf p-k}}-
\omega_{\nu{\bf p}})]\}, \\
A_{T}({\bf k},\omega)&=&\pi{1\over 4N}\sum_{\bf p}\sum_{\mu\nu}
(-1)^{\mu+\nu}Z_{hFA}^{(\mu)}{B_{\nu{\bf p}}\over\omega_{\nu{\bf p}}
}\{ L^{(1)}_{\mu\nu}({\bf k,p})[U^{2}_{h\mu{\bf p-k}}
\delta(\omega+E_{h\mu{\bf p-k}}-\omega_{\nu{\bf p}})+V^{2}_{h\mu{\bf
p-k}}\delta(\omega-E_{h\mu{\bf p-k}}+ \omega_{\nu{\bf p}})]\nonumber\\
&+&L^{(2)}_{\mu\nu}({\bf k,p})[U^{2}_{h\mu{\bf p-k}}
\delta(\omega+E_{h\mu{\bf p-k}}+ \omega_{\nu{\bf p}})
+V^{2}_{h\mu{\bf p-k}}\delta(\omega-E_{h\mu{\bf p-k}}-
\omega_{\nu{\bf p}})]\},
\end{eqnarray}
\begin{eqnarray}
\Delta_{L}(\bf k)&=&-{1\over 16N}\sum_{{\bf p},\mu,\nu}
Z^{(\mu)}_{hFA} {\bar{\Delta}_{hz}^{(\mu)}({\bf p-k}) \over
E_{h\mu{\bf p-k}}}{B_{\nu{\bf p}}\over\omega_{\nu{\bf p}}}{\rm th}
[{1\over 2}\beta E_{h\mu{\bf p-k}}]{\rm coth}
[{1\over 2}\beta\omega_{\nu{\bf p}}], \\
\Delta_{T}(\bf k)&=&-{1\over 16N}\sum_{{\bf p},\mu,\nu}
(-1)^{\mu+\nu}Z^{(\mu)}_{hFA}{\bar{\Delta}_{hz}^{(\mu)}({\bf p-k})
\over E_{h\mu{\bf p-k}}}{B_{\nu{\bf p}}\over\omega_{\nu{\bf p}}}
{\rm th}[{1\over 2}\beta E_{h\mu{\bf p-k}}]{\rm coth} [{1\over 2}
\beta\omega_{\nu{\bf p}}].
\end{eqnarray}
\end{subequations}
\end{widetext}
With the above longitudinal and transverse parts of the SC gap
functions in Eqs. (10c) and (10d), the corresponding longitudinal
and transverse SC gap parameters are obtained as $\Delta_{L}=-
\chi_{1}\Delta_{hL}$ and $\Delta_{T}=-\chi_{\perp}\Delta_{hT}$,
respectively. In the bilayer coupling case, the more appropriate
classification is in terms of the spectral function and SC gap
function within the basis of the antibonding and bonding components
\cite{kordyuk,dfeng,chuang,borisenko,dfeng1}. In this case, the
electron spectral function and SC gap parameter can be transformed
from the plane representation to the antibonding-bonding
representation as,
\begin{subequations}
\begin{eqnarray}
A^{(a)}({\bf k},\omega)&=&{1\over 2}[A_{L}({\bf k},\omega)-A_{T}
({\bf k},\omega)],\\
A^{(b)}({\bf k},\omega)&=&{1\over 2}[A_{L}({\bf k},\omega)+ A_{T}
({\bf k},\omega)],\\
\Delta^{(a)}&=&\Delta_{L}-\Delta_{T}, \\
\Delta^{(b)}&=&\Delta_{L}+\Delta_{T}.
\end{eqnarray}
\end{subequations}
respectively, then the antibonding and bonding parts have odd and
even symmetries, respectively.

\section{Electron structure of bilayer cuprate superconductors}

We now begin to discuss the effect of the bilayer interaction on the
electronic structure in the SC state. We first plot, in Fig. 1, the
antibonding (solid line) and bonding (dashed line) electron spectral
functions in the $[\pi,0]$ point for parameters $t/J=2.5$,
$t'/t=0.3$, and $t_{\perp}/t=0.35$ with temperature $T=0.002J$ at
the doping concentration $\delta=0.15$. In comparison with the
single layer case \cite{guo}, the electron spectrum of the bilayer
system has been split into the bonding and antibonging components,
with the bonding and antibonding SC quasiparticle peaks in the
$[\pi,0]$ point are located at the different positions. In this
sense, the differentiation between the bonding and antibonding
components of the electron spectral function is essential. The
antibonding spectrum consists of a low energy antibonding peak,
corresponding to the SC peak, and the bonding spectrum has a higher
energy bonding peak, corresponding to the hump, while the spectral
dip is in between them, then the total contributions for the
electron spectrum from both antibonding and bonding components give
rise to the PDH structure. Although the simple bilayer $t$-$t'$-$J$
model (1) cannot be regarded as a comprehensive model for a
quantitative comparison with bilayer cuprate superconductors, our
present results for the SC state are in qualitative agreement with
the major experimental observations on bilayer cuprate
superconductors
\cite{shen,dessau,randeria,fedorov,lu,sato,borisenko}.

\begin{figure}
\includegraphics[scale=0.52]{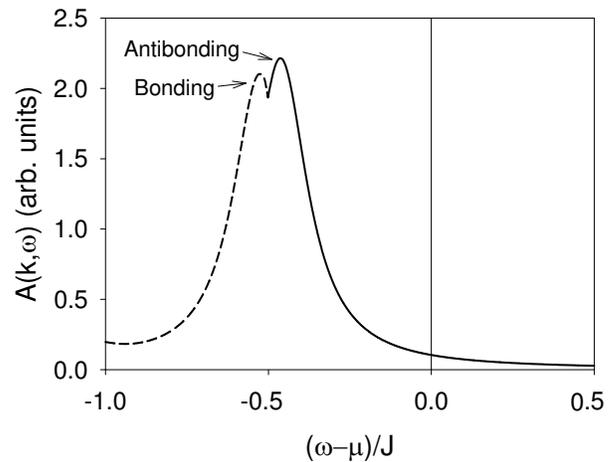}
\caption{The antibonding (solid line) and bonding (dashed line)
electron spectral functions in the $[\pi,0]$ point for $t/J=2.5$,
$t'/t=0.3$, and $t_{\perp}/t=0.35$ with $T=0.002J$ at
$\delta=0.15$.}
\end{figure}

\begin{figure}
\includegraphics[scale=0.52]{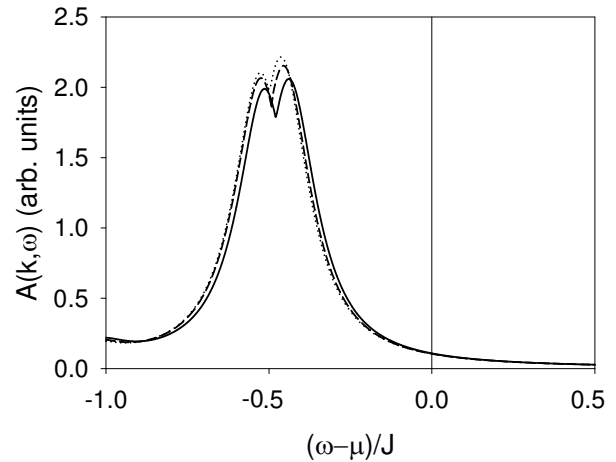}
\caption{The electron spectral functions at $[\pi,0]$ point for
$t/J=2.5$, $t'/t=0.3$, and $t_{\perp}/t=0.35$ with $T=0.002J$ at
$\delta=0.09$ (solid line), $\delta=0.12$ (dashed line), and
$\delta=0.15$ (dotted line).}
\end{figure}

We now turn to discuss the doping evolution of the electron spectrum
of bilayer cuprate superconductors in the SC state. We have
calculated the electron spectrum at different doping concentrations,
and the result of the electron spectral functions in the $[\pi,0]$
point for $t/J=2.5$, $t'/t=0.3$, and $t_{\perp}/t=0.35$ with
$T=0.002J$ at $\delta=0.09$ (solid line), $\delta=0.12$ (dashed
line), and $\delta=0.15$ (dotted line) are plotted in Fig. 2. In
comparison with the corresponding ARPES experimental results of the
bilayer cuprate superconductor
Bi$_{2}$Sr$_{2}$CaCu$_{2}$O$_{8+\delta}$ in the SC state in Ref.
\cite{campuzano2}, it is obviously that the doping evolution of the
spectral weight of the bilayer superconductor
Bi$_{2}$Sr$_{2}$CaCu$_{2}$O$_{8+\delta}$ is reproduced. With
increasing the doping concentration, both SC peak and hump become
sharper, and then the spectral weights increase in intensity.
Furthermore, we have also calculated the electron spectrum with
different temperatures, and the results show that the spectral
weights of both SC peak and hump are suppressed with increasing
temperatures. Our these results are also qualitatively consistent
with the ARPES experimental results on bilayer cuprate
superconductors in the SC state \cite{shen,fedorov,campuzano2}.

To better perceive the anomalous form of the antibonding and bonding
electron spectral functions as a function of energy $\omega$ for
${\bf k}$ in the vicinity of the $[\pi,0]$ point, we have made a
series of calculations for the electron spectral function at
different momenta, and the results show that the sharp SC peak from
the electron antibonding spectral function and hump from the bonding
spectral function persist in a very large momentum space region
around the $[\pi,0]$ point. To show this point clearly, we plot the
positions of the antibonding peak and bonding hump in the electron
spectrum as a function of momentum along the direction
$[-0.2\pi,\pi]\rightarrow [0,\pi]\rightarrow [0.2\pi,\pi]$ with
$T=0.002J$ at $\delta=0.15$ for $t/J=2.5$, $t'/t=0.3$, and
$t_{\perp}/t=0.35$ in Fig. 3. Our result shows that there are two
branches in the quasiparticle dispersion, with upper branch
corresponding to the antibonding quasiparticle dispersion, and lower
branch corresponding to the bonding quasiparticle dispersion.
Furthermore, the BS reaches its maximum at the $[\pi,0]$ point. Our
present result also shows that in analogy to the two flat bands
appeared in the normal state \cite{lan}, both electron antibonding
peak and bonding hump have a weak dispersion around the $[\pi,0]$
point, in qualitative agreement with the ARPES experimental
measurements on bilayer cuprate superconductors in the SC state
\cite{shen,dessau,randeria,fedorov,lu,sato,dfeng}.

\begin{figure}
\includegraphics[scale=0.52]{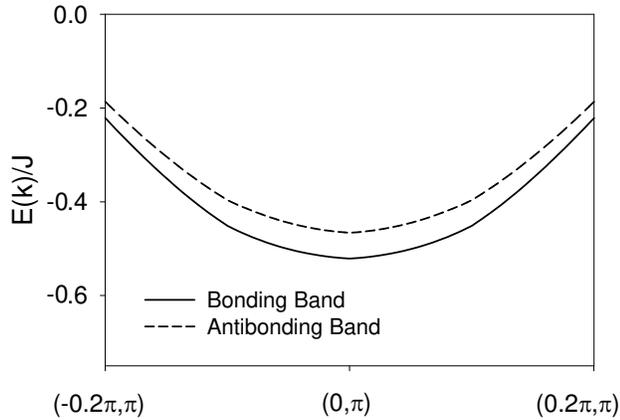}
\caption{The positions of the antibonding peaks and bonding humps in
the electron spectrum as a function of momentum along the direction
$[-0.2\pi,\pi]\rightarrow [0,\pi]\rightarrow [0.2\pi,\pi]$ with
$T=0.002J$ at $\delta=0.15$ for $t/J=2.5$, $t'/t=0.3$, and
$t_{\perp}/t=0.35$.}
\end{figure}

\begin{figure}
\includegraphics[scale=0.52]{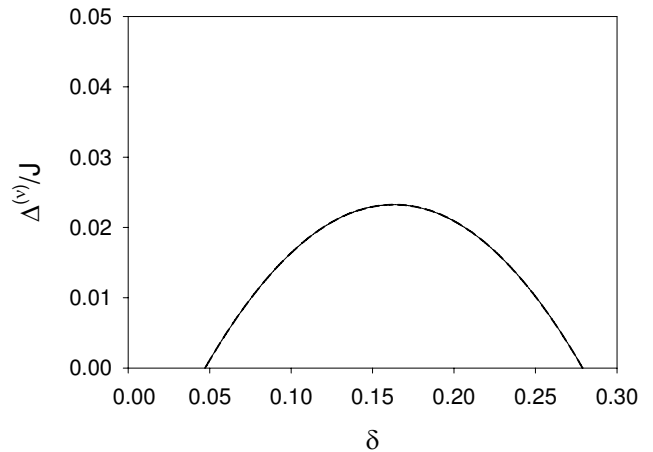}
\caption{The antibonding (solid line) and bonding (dashed line) gap
parameters as a function of the doping concentration with $T=0.002J$
for $t/J=2.5$, $t'/t=0.3$, and $t_{\perp}/t=0.35$.}
\end{figure}

In the above calculations, we find that although the antibonding SC
peak and bonding hump have different dispersions, the transverse
part of the SC gap parameter $\Delta_{T}\approx 0$. To show this
point clearly, we plot the antibonding and bonding gap parameters in
Eqs. (11c) and (11d) as a function of the doping concentration with
$T=0.002J$ for $t/J=2.5$, $t'/t=0.3$, and $t_{\perp}/t=0.35$ in Fig.
4. As seen from Fig. 4, both antibonding and bonding gap parameters
have the same d-wave SC gap magnitude in a given doping
concentration, i.e., $\Delta^{a}\approx\Delta^{b}$. This result
shows that although there is a single electron interlayer coherent
hopping (2) in bilayer cuprate superconductors in the SC state, the
electron interlayer pairing interaction vanishes. This reflects that
in the present kinetic energy driven SC mechanism, the weak dressed
holon-spin interaction due to the interlayer coherent hopping (2)
from the kinetic energy terms in Eq. (3) does not induce the dressed
holon interlayer pairing state by exchanging spin excitations in the
higher power of the doping concentration. This is different from the
dressed holon-spin interaction due to the intralayer hopping from
the kinetic energy terms in Eq. (3), it can induce superconductivity
by exchanging spin excitations in the higher power of the doping
concentration \cite{feng}. Our this result is also consistent with
the ARPES experimental results of the bilayer cuprate superconductor
Bi(Pb)$_{2}$Sr$_{2}$CaCu$_{2}$O$_{8+\delta}$ \cite{dfeng,borisenko},
where the SC gap separately for the bonding and antibonding bands
has been measured, and it is found that both d-wave SC gaps from the
antibonding and bonding components are identical within the
experimental uncertainties.

To our present understanding, two main reasons why the electronic
structure of bilayer cuprate superconductors in the SC state can be
described qualitatively in the framework of the kinetic energy
driven superconductivity by considering the bilayer interaction are
as follows. Firstly, the bilayer interaction causes the BS, this
leads to that the full electron normal (anomalous) Green's function
is divided into the longitudinal and transverse parts, respectively,
then the bonding and antibonding electron spectral functions (SC gap
functions) are obtained from these longitudinal and transverse parts
of the electron normal (anomalous) Green's function, respectively.
Although the transverse part of the SC gap parameter
$\Delta_{T}\approx 0$, the antibonding peak around the $[\pi,0]$
point is always at lower binding energy than the bonding peak (hump)
due to the BS. In this sense, the PDH structure in the bilayer
cuprate superconductors in the SC state is mainly caused by the BS.
Secondly, the SC state in the kinetic energy driven SC mechanism is
the conventional BCS like as in the single layer case
\cite{guo,feng}. This can be understood from the electron normal and
anomalous Green's functions in Eq. (9). Since the spins center
around the $[\pi,\pi]$ point in the MF level \cite{guo,feng,lan},
then the main contributions for the spins comes from the $[\pi,\pi]$
point. In this case, the longitudinal and transverse parts of the
electron normal and anomalous Green's functions in Eq. (9) can be
approximately reduced in terms of $\omega_{\nu{\bf p}=[\pi,\pi]}\sim
0$ and one of the self-consistent equations \cite{lan} $1/2=\langle
S_{ia}^{+}S_{ia}^{-}\rangle= 1/(4N)\sum_{\nu,{\bf k}}(B_{\nu{\bf k}}
/\omega_{\nu{\bf k}}){\rm coth} [(1/2)\beta\omega_{\nu{\bf k}}]$ as,
\begin{subequations}
\begin{align}
G_{L}({\bf k},\omega)&\approx{1\over 2}\sum_{\nu=1,2}Z^{(\nu)}_{FA}
\left ({U^{2}_{\nu{\bf k}}\over\omega-E_{\nu{\bf k}}}+
{V^{2}_{\nu{\bf k}}\over\omega+E_{\nu{\bf k}}}\right), \\
G_{T}({\bf k},\omega)&\approx{1\over
2}\sum_{\nu=1,2}(-1)^{\nu+1}Z^{(\nu)}_{FA}\left ({U^{2}_{\nu{\bf
k}}\over\omega-E_{\nu{\bf k}}} +{V^{2}_{\nu{\bf
k}}\over\omega+E_{\nu{\bf k}}}\right ),\\
\Gamma^{\dagger}_{L}({\bf k},\omega)&={1\over 2}\sum_{\nu=1,2}
Z^{(\nu)}_{FA}{\bar{\Delta}_{z}^{(\nu)}({\bf k})\over 2E_{\nu{\bf
k}}}\left ({1\over\omega-E_{\nu{\bf k}}}-{1\over\omega+E_{\nu{\bf
k}}}\right ),\\
\Gamma^{\dagger}_{T}({\bf k},\omega)&={1\over 2}\sum_{\nu=1,2}
(-1)^{\nu+1}Z^{(\nu)}_{FA}{\bar{\Delta}_{z}^{(\nu)}({\bf k})
\over 2E_{\nu{\bf k}}}\nonumber\\
&\qquad\times\left ({1\over\omega-E_{\nu{\bf k}}}-{1\over\omega
+E_{\nu{\bf k}}}\right ),
\end{align}
\end{subequations}
where the electron coherent weights $Z^{(\nu)}_{FA}=Z^{(\nu)}_{hFA}
/2$, the electron quasiparticle coherence factors $U^{2}_{\nu{\bf k}
}\approx V^{2}_{h\nu{\bf k}-{\bf k}_{A}}$ and $V^{2}_{\nu{\bf k}}
\approx U^{2}_{h\nu{\bf k}-{\bf k}_{A}}$, the SC gap function
${\bar{\Delta}_{z}^{(\nu)}({\bf k})}\approx
{\bar{\Delta}_{hz}^{(\nu)}({\bf k}-{\bf k}_{A})}$ and the electron
quasiparticle spectrum $E_{\nu{\bf k}}\approx E_{h\nu{\bf k}-{\bf
k_{A}}}$, with ${\bf k_{A}}= [\pi,\pi]$. As in the single layer case
\cite{guo,feng}, this reflects that the hole-like dressed holon
quasiparticle coherence factors $V_{h\nu{\bf k}}$ and $U_{h\nu{\bf
k}}$ and hole-like dressed holon quasiparticle spectrum $E_{h\nu{\bf
k}}$ have been transferred into the electron quasiparticle coherence
factors $U_{\nu{\bf k}}$ and $V_{\nu{\bf k}}$ and electron
quasiparticle spectrum $E_{\nu{\bf k}}$, respectively, by the
convolutions of the corresponding longitudinal and transverse parts
of the MF spin Green's function and full dressed holon normal and
anomalous Green's functions due to the charge-spin recombination
\cite{anderson}. As a result, these electron normal and anomalous
Green's functions in Eq. (12) are typical bilayer BCS like
\cite{bcs}. This also reflects that as in the single layer case
\cite{guo,feng}, the dressed holon pairs condense with the d-wave
symmetry in a wide range of the doping concentration, then the
electron Cooper pairs originating from the dressed holon pairing
state are due to the charge-spin recombination, and their
condensation automatically gives the electron quasiparticle
character. These are why the basic bilayer BCS formalism \cite{bcs}
is still valid in discussions of SC coherence of the quasiparticle
peak and hump, although the pairing mechanism is driven by the
intralayer kinetic energy by exchanging spin excitations, and other
exotic magnetic scattering \cite{dai,arai} is beyond the BCS
formalism.

\section{Summary and discussions}

We have studied the electronic structure of bilayer cuprate
superconductors in the SC state based on the kinetic energy driven
SC mechanism \cite{feng}. Our results show that the electron
spectrum of bilayer cuprate superconductors is split into the
bonding and antibonding components by the BS, then the observed PDH
structure around the $[\pi,0]$ point is mainly caused by this BS,
with the SC peak being related to the antibonding component, and the
hump being formed by the bonding component. The spectral weight
increases with increasing the doping concentration. In analogy to
the two flat bands appeared in the normal state, the antibonding and
bonding quasiparticles around the $[\pi,0]$ point disperse weakly
with momentum, in qualitative agreement with the experimental
observation on the bilayer cuprate superconductors
\cite{shen,dessau,randeria,fedorov,lu,sato}. Our these results also
show that the bilayer interaction has significant contributions to
the electronic structure of bilayer cuprate superconductors in the
SC state.

It has been shown from the ARPES experiments \cite{dfeng,shen} that
the BS has been detected in both normal and SC states, and then the
electron spectral functions display the double-peak structure in the
normal state and PDH structure in the SC state. Recently, we
\cite{lan} have studied the electron spectrum of bilayer cuprate
superconductors in the normal state, and shown that the double-peak
structure in the electron spectrum in the normal state is dominated
by the BS. On the other hand, although the antibonding and bonding
SC peaks have different dispersions, the antibonding and bonding
parts have the same d-wave SC gap amplitude as mentioned above.
Incorporating our previous discussions for the normal state case
\cite{lan} and the present studies for the SC state case, we
therefore find that the one of the important roles of the interlayer
coherent hopping (2) is to split the electron spectrum of the
bilayer system into the bonding and antibonding components in both
normal and SC states. As a consequence, the well pronounced PDH
structure of bilayer cuprate superconductors in the SC state and
double-peak structure in the normal state are mainly caused by the
BS.

\acknowledgments The authors would like to thank Dr. H. Guo and Dr.
L. Cheng for the helpful discussions. This work was supported by the
National Natural Science Foundation of China under Grant No.
90403005, and the funds from the Ministry of Science and Technology
of China under Grant Nos. 2006CB601002 and 2006CB921300.

\appendix

\section{Dressed holon BCS type normal and anomalous Green's
functions in bilayer cuprate superconductors}

In the single layer case, it has been shown \cite{feng} that the
dressed holon-spin interactions from the kinetic energy terms of the
$t$-$t'$-$J$ model are quite strong, and in the case without AFLRO,
these interactions can induce the dressed holon pairing state (then
the electron Cooper pairing state) by exchanging spin excitations in
the higher power of the doping concentration. Following their
discussions \cite{guo,feng}, we obtain in terms of Eliashberg's
strong coupling theory \cite{eliashberg} that the self-consistent
equations that satisfied by the full dressed holon normal and
anomalous Green's functions in the bilayer system in the SC state
as,
\begin{subequations}
\begin{eqnarray}
g({\bf k},\omega)&=&g^{(0)}({\bf k},\omega)+g^{(0)}({\bf k},
\omega)\left[\Sigma^{(h)}_{1}({\bf k},\omega)g({\bf k},\omega)\right.\nonumber\\
&&\qquad\qquad-\left.\Sigma^{(h)}_{2}(-{\bf
k},-\omega)\Im^{\dagger}({\bf
k},\omega)\right], \\
\Im^{\dagger}({\bf k},\omega)&=&g^{(0)}(-{\bf k},-\omega)
\left[\Sigma^{(h)}_{1}(-{\bf k},-\omega)\Im^{\dagger}(-{\bf
k},-\omega)\right.\nonumber\\
&&\qquad\qquad +\left.\Sigma^{(h)}_{2}(-{\bf k},-\omega)g({\bf
k},\omega)\right],
\end{eqnarray}
\end{subequations}
respectively, where the MF dressed holon normal Green's function
\cite{lan} $g^{(0)}({\bf k},\omega)=g^{(0)}_{L}({\bf k},\omega)+
\sigma_{x}g^{(0)}_{T}({\bf k},\omega)$, with the longitudinal and
transverse parts are evaluated as $g^{(0)}_{L}({\bf k},\omega)=(1/2)
\sum_{\nu=1,2} (\omega-\xi_{\nu{\bf k}})^{-1}$ and $g^{(0)}_{T}
({\bf k},\omega)=(1/2)\sum_{\nu=1,2}(-1)^{\nu+1} (\omega-
\xi_{\nu{\bf k}})^{-1}$, respectively, while the dressed holon
self-energy functions $\Sigma^{(h)}_{1}({\bf k},\omega)=
\Sigma^{(h)}_{1L}({\bf k},\omega) + \sigma_{x}\Sigma^{(h)}_{1T}
({\bf k},\omega)$ and $\Sigma^{(h)}_{2}({\bf k},\omega)=
\Sigma^{(h)}_{2L}({\bf k},\omega) + \sigma_{x} \Sigma^{(h)}_{2T}
({\bf k},\omega)$, with the corresponding longitudinal and
transverse parts have been given in Eq. (6).

In the previous discussions of the electronic structure for the
single layer cuprate superconductors in the SC state \cite{guo}, it
has been shown the self-energy function $\Sigma^{(h)}_{2}({\bf k},
\omega)$ describes the effective dressed holon pair gap function,
while the self-energy function $\Sigma^{(h)}_{1}({\bf k},\omega)$
describes the quasiparticle coherence. Since $\Sigma^{(h)}_{2}({\bf
k},\omega) $ is an even function of $\omega$, while
$\Sigma^{(h)}_{1}({\bf k}, \omega)$ is not, therefore for the
convenience, the self-energy function $\Sigma^{(h)}_{1}({\bf k},
\omega)$ can be broken up into its symmetric and antisymmetric parts
as, $\Sigma^{(h)}_{1}({\bf k}, \omega)=\Sigma^{(h)}_{1e}({\bf
k},\omega)+\omega \Sigma^{(h)}_{1o} ({\bf k},\omega)$, then both
$\Sigma^{(h)}_{1e} ({\bf k},\omega)$ and $\Sigma^{(h)}_{1o}({\bf k},
\omega)$ are even functions of $\omega$. Now we can define the
dressed holon quasiparticle coherent weights in the present bilayer
system as $Z^{-1}_{hF1}({\bf k}, \omega)=1-\Sigma^{(ho)}_{1L}({\bf
k},\omega)$ and $Z^{-1}_{hF2} ({\bf k},\omega)=\Sigma^{(ho)}_{1T}
({\bf k},\omega)$. As in the single layer case \cite{guo}, we only
discuss the low-energy behavior of the electronic structure of
bilayer cuprate superconductors, which means that the effective
dressed holon pair gap functions and quasiparticle coherent weights
can be discussed in the static limit,  i.e., $\bar{\Delta}_{h} ({\bf
k})=\Sigma^{(h)}_{2}({\bf k},\omega)\mid_{\omega=0}=
\bar{\Delta}_{hL}({\bf k})+\sigma_{x}\bar{\Delta}_{hT}({\bf k})$,
$Z^{-1}_{hF1}({\bf k})=1-\Sigma^{(ho)}_{1L}({\bf k},\omega)
\mid_{\omega=0}$ and $Z^{-1}_{hF2}({\bf k})= \Sigma^{(ho)}_{1T}
({\bf k},\omega)\mid_{\omega=0}$. As in the single layer case
\cite{guo}, although $Z_{hF1}({\bf k})$ and $Z_{hF2}({\bf k})$ still
are a function of ${\bf k}$, the wave vector dependence may be
unimportant. This followed from the ARPES experiments \cite{shen}
that in the SC-state of bilayer cuprate superconductors, the lowest
energy states are located at the $[\pi,0]$ point, which indicates
that the majority contribution for the electron spectrum comes from
the $[\pi,0]$ point. In this case, the wave vector ${\bf k}$ in
$Z_{hF1}({\bf k})$ and $Z_{hF2}({\bf k} )$ can be chosen as
$Z^{-1}_{hF1}=1-\Sigma^{(ho)}_{1L}({\bf k}) \mid_{{\bf k}=[\pi,0]}$
and $Z^{-1}_{hF2}=\Sigma^{(ho)}_{1T}({\bf k} )\mid_{{\bf k}=
[\pi,0]}$. With the help of the above discussions, the corresponding
longitudinal and transverse parts of the dressed holon normal and
anomalous Green's functions in Eqs. (A1a) and (A1b) now can be
obtained explicitly as,
\begin{widetext}
\begin{subequations}
\begin{eqnarray}
g_{L}({\bf k},\omega) &=& {1\over 2}\sum_{\nu=1,2}Z^{(\nu)}_{hFA}
\left ({U^{2}_{h\nu{\bf k}}\over\omega-E_{h\nu{\bf k}}}+
{V^{2}_{h\nu{\bf k}}\over\omega+E_{h\nu{\bf k}}}\right), \\
g_{T}({\bf k},\omega)&=&{1\over 2}\sum_{\nu=1,2}(-1)^{\nu+1}
Z^{(\nu)}_{hFA}\left ({U^{2}_{h\nu{\bf k}}\over\omega- E_{h\nu{\bf
k}}}+{V^{2}_{h\nu{\bf k}}\over\omega+E_{h\nu{\bf k}}}
\right ), \\
\Im^{\dagger}_{L}({\bf k},\omega)&=&-{1\over 2}\sum_{\nu=1,2}
Z^{(\nu)}_{hFA}{\bar{\Delta}_{hz}^{(\nu)}({\bf k})\over 2E_{h\nu{\bf
k}}}\left ({1\over \omega-E_{h\nu{\bf k}}}-{1\over
\omega+ E_{h\nu{\bf k}}}\right ), \\
\Im^{\dagger}_{T}({\bf k},\omega)&=&-{1\over 2}\sum_{\nu=1,2}
(-1)^{\nu+1}Z^{(\nu)}_{hFA}{\bar{\Delta}_{hz}^{(\nu)}({\bf k})\over
2E_{h\nu{\bf k}}}\left ({1\over\omega-E_{h\nu{\bf k}}}-{1\over
\omega+E_{h\nu{\bf k}}}\right ),
\end{eqnarray}
\end{subequations}
\end{widetext}
with the dressed holon effective gap parameters and quasiparticle
coherent weights satisfy the following four equations,
\begin{widetext}
\begin{subequations}
\begin{eqnarray}
\bar{\Delta}_{hL}&=&-{4\over 32N^{3}}\sum_{{\bf k,q,p}}
\sum_{\nu,\nu',\nu''}\gamma^{(d)}_{\bf k-p+q}C_{\nu\nu''}({\bf k+q})
{Z^{(\nu'')}_{hFA}B_{\nu'{\bf p}} B_{\nu{\bf q}}\over
\omega_{\nu'{\bf p}}\omega_{\nu{\bf q}}}\bar{\Delta}^{(\nu'')}_{hz}
({\bf k})\nonumber \\
&\times&\left ({F^{(1)}_{\nu\nu'\nu''}({\bf q,p})+
F^{(2)}_{\nu\nu'\nu''}({\bf k,q,p})\over [\omega_{\nu'{\bf p}}
-\omega_{\nu{\bf q}}]^2-E^{2}_{h\nu''{\bf k}}}+
{F^{(3)}_{\nu\nu'\nu''}({\bf q,p})+F^{(4)}_{\nu\nu'\nu''}({\bf k, q,
p})\over [\omega_{\nu'{\bf p}}+\omega_{\nu{\bf q}}]^{2} -
E^{2}_{h\nu''{\bf k}}} \right ),\\
\bar{\Delta}_{hT}&=&-{1\over
32N^{3}}\sum_{{\bf k,q,p}}
\sum_{\nu,\nu',\nu''}(-1)^{\nu+\nu'+\nu''+1}C_{\nu\nu''}({\bf k+q})
{Z^{(\nu'')}_{hFA}B_{\nu'{\bf p}} B_{\nu{\bf q}}\over
\omega_{\nu'{\bf p}}\omega_{\nu{\bf q}}}\bar{\Delta}^{(\nu'')}_{hz}
({\bf k})\nonumber \\
&\times&\left ({F^{(1)}_{\nu\nu'\nu''}({\bf q,p})+
F^{(2)}_{\nu\nu'\nu''}({\bf k,q,p})\over [\omega_{\nu'{\bf p}}
-\omega_{\nu{\bf q}}]^2-E^{2}_{h\nu''{\bf k}}}+
{F^{(3)}_{\nu\nu'\nu''}({\bf q,p})+F^{(4)}_{\nu\nu'\nu''}({\bf k, q,
p})\over [\omega_{\nu'{\bf p}}+\omega_{\nu{\bf q}}]^{2} -
E^{2}_{h\nu''{\bf k}}} \right ),\\
{1\over Z_{hFA}^{(1)}}&=&1+{1\over 32N^{2}}\sum_{{\bf q,p}}
\sum_{\nu,\nu',\nu''}[1+(-1)^{\nu+\nu'+\nu''+1}]C_{\nu\nu''}({\bf p+
k_{0}}){Z^{(\nu'')}_{hFA} B_{\nu'{\bf p}}B_{\nu{\bf q}}\over
\omega_{\nu'{\bf p}} \omega_{\nu{\bf q}}}\nonumber\\
&\times&\left({H^{(1)}_{\nu\nu'\nu''}({\bf q,p})\over
[\omega_{\nu'{\bf p} }-\omega_{\nu{\bf q}}+E_{h\nu''{\bf p-q+
k_{0}}}]^{2}}+{H^{(2)}_{\nu\nu'\nu''}({\bf q,p})\over
[\omega_{\nu'{\bf p}} -\omega_{\nu{\bf q}}-E_{h\nu''{\bf p-q+
k_{0}}}]^{2}}\right. \nonumber\\
&+&\left.{H^{(3)}_{\nu\nu'\nu''}({\bf q,p})\over[\omega_{\nu'{\bf p}
}+\omega_{\nu{\bf q}}+E_{h\nu''{\bf p-q+k_{0}}}]^{2}}
+{H^{(4)}_{\nu\nu'\nu''}({\bf q,p})\over [\omega_{\nu' {\bf p}}
+\omega_{\nu{\bf q}}-E_{h\nu''{\bf p-q+k_{0}}}]^{2}}\right), \\
{1\over Z_{hFA}^{(2)}}&=&1+{1\over 32N^{2}}\sum_{{\bf q,p}}
\sum_{\nu,\nu',\nu''}[1-(-1)^{\nu+\nu'+\nu''+1}]C_{\nu\nu''}({\bf p+
k_{0}}){Z^{(\nu'')}_{hFA} B_{\nu'{\bf p}}B_{\nu{\bf q}}\over
\omega_{\nu'{\bf p}} \omega_{\nu{\bf q}}}\nonumber\\
&\times&\left({H^{(1)}_{\nu\nu'\nu''}({\bf q,p})\over
[\omega_{\nu'{\bf p} }-\omega_{\nu{\bf q}}+E_{h\nu''{\bf p-q+
k_{0}}}]^{2}}+{H^{(2)}_{\nu\nu'\nu''}({\bf q,p})\over
[\omega_{\nu'{\bf p}} -\omega_{\nu{\bf q}}-E_{h\nu''{\bf p-q+
k_{0}}}]^{2}}\right. \nonumber\\
&+&\left.{H^{(3)}_{\nu\nu'\nu''}({\bf q,p})\over[\omega_{\nu'{\bf p}
}+\omega_{\nu{\bf q}}+E_{h\nu''{\bf p-q+k_{0}}}]^{2}}
+{H^{(4)}_{\nu\nu'\nu''}({\bf q,p})\over [\omega_{\nu' {\bf p}}
+\omega_{\nu{\bf q}}-E_{h\nu''{\bf p-q+k_{0}}}]^{2}}\right),
\end{eqnarray}
\end{subequations}
\end{widetext}
where $C_{\nu\nu''}({\bf k})=[Z(t\gamma_{\bf k}-t'\gamma{'}_{\bf
k}) +(-1)^{\nu+\nu''}t_{\perp}({\bf k})]^{2}$,
$F^{(1)}_{\nu\nu'\nu''} ({\bf q, p})= n_{B} (\omega_{\nu{\bf
q}})+n_{B}(\omega_{\nu'{\bf p}} )+2n_{B} (\omega_{\nu{\bf
q}})n_{B}(\omega_{\nu'{\bf p}})$, $F^{(2)}_{\nu\nu'\nu''}({\bf
k,q,p})=[2n_{F}(E_{h\nu''{\bf k}}) -1] [\omega_{\nu'{\bf
p}}-\omega_{\nu{\bf q}}][n_{B}(\omega_{\nu{\bf q}}
)-n_{B}(\omega_{\nu'{\bf p}})]/E_{h\nu''{\bf k}}$,
$F^{(3)}_{\nu\nu'\nu''}({\bf q,p})=1+n_{B}(\omega_{\nu{\bf q}})+
n_{B}(\omega_{\nu'{\bf p}})+2n_{B}(\omega_{\nu{\bf q}}) n_{B}
(\omega_{\nu'{\bf p}})$, $F^{(4)}_{\nu\nu'\nu''}({\bf k,q,p})=
[2n_{F}(E_{h\nu''{\bf k}})-1] [\omega_{\nu'{\bf p}} +
\omega_{\nu{\bf q}}][1+n_{B}(\omega_{\nu{\bf q}})+ n_{B}
(\omega_{\nu'{\bf p}})]/E_{h\nu''{\bf k}}$,
$H^{(1)}_{\nu\nu'\nu''}({\bf q,p})=n_{F}(E_{h\nu''{\bf p-q+k_{0}}})
[n_{B}(\omega_{\nu'{\bf p}})-n_{B} (\omega_{\nu{\bf q}})]+
n_{B}(\omega_{\nu{\bf q}})[1+ n_{B} (\omega_{\nu'{\bf p}})]$,
$H^{(2)}_{\nu\nu'\nu''}({\bf q,p})=n_{F}(E_{h\nu''{\bf p-q+k_{0}}})
[n_{B}(\omega_{\nu{\bf q}})-n_{B} (\omega_{\nu'{\bf p}})]+
n_{B}(\omega_{\nu'{\bf p}})[1+n_{B} (\omega_{\nu{\bf q}})]$,
$H^{(3)}_{\nu\nu'\nu''}({\bf q, p})=[1-n_{F}(E_{h\nu''{\bf
p-q+k_{0}}})][1+n_{B}(\omega_{\nu{\bf q}}) +n_{B}(\omega_{\nu'{\bf
p}})]+n_{B}(\omega_{\nu{\bf q}})n_{B} (\omega_{\nu'{\bf p}})$,
$H^{(4)}_{\nu\nu'\nu''}({\bf q,p})=n_{F}(E_{h\nu''{\bf p-q+k_{0}}})
[1+n_{B}(\omega_{\nu{\bf q}})+ n_{B}(\omega_{\nu'{\bf p}})]+
n_{B}(\omega_{\nu{\bf q}}) n_{B} (\omega_{\nu'{\bf p}})$, and ${\bf
k}_{0}=[\pi,0]$. These four equations must be solved
self-consistently in combination with other equations as in the
single layer case \cite{guo,feng}, then all order parameters,
decoupling parameter $\alpha$, and chemical potential $\mu$ are
determined by the self-consistent calculation.

\end{document}